\documentclass[11pt]{cernrep}
\usepackage{graphicx}
\usepackage{here}

\begin{document}
\bibliographystyle{unsrt}
\renewcommand{\thefootnote}{\fnsymbol{footnote}}

\begin{flushright}
hep-ph/0401061\\
LU TP 04--05\\
January 2004
\end{flushright}

\vspace{10mm}

\noindent{\Large\bf NEW SHOWERS WITH TRANSVERSE-MOMENTUM-ORDERING%
\footnote{submitted to the proceedings of the Workshop on 
Physics at TeV Colliders, Les Houches, France, 
26 May -- 6 June 2003}}\\[8mm]
\textit{T. Sj\"ostrand}\\
Department of Theoretical Physics, Lund University, 
S\"olvegatan 14A, S-223 62 Lund, Sweden

\section{INTRODUCTION}

The initial- \cite{Sjostrand:1985xi,Bengtsson:1986gz,Miu:1998ju} 
and final-state \cite{Bengtsson:1987et,Norrbin:2000uu} showers 
in the \textsc{Pythia} event generator 
\cite{Sjostrand:2000wi,Sjostrand:2003wg} are based on 
virtuality-ordering, i.e.\ uses spacelike $Q^2$ and timelike $M^2$, 
respectively, as evolution variables. Other algorithms in common 
use are the angular-ordered ones in \textsc{Herwig} 
\cite{Marchesini:1984bm,Corcella:2000bw} and the 
$p_{\perp}$-ordered dipole-based ones in \textsc{Ariadne/Ldc}
\cite{Lonnblad:1992tz,Kharraziha:1998dn}. All three have been 
comparably successful, in terms of ability to predict or describe 
data, and therefore have offered useful cross-checks. Some 
shortcomings of the virtuality-ordering approach, with respect to
coherence conditions, have been compensated (especially relative to
\textsc{Herwig}) by a better coverage of phase space and more 
efficient possibilities to merge smoothly with first-order matrix 
elements.

Recently, the possibility to combine matrix elements of several 
orders consistently with showers has been raised 
\cite{Catani:2001cc,Lonnblad:2001iq}, e.g.\ $\mathrm{W} + n$ jets,
$n= 0, 1, 2, 3, \ldots$. In such cases, a $p_{\perp}$-ordering 
presumably offers the best chance to provide a sensible definition 
of hardness. It may also tie in better e.g.\ with the 
$p_{\perp}$-ordered approach to multiple interactions 
\cite{multinthere}. This note therefore is a study of
how the existing \textsc{Pythia} algorithms can be reformulated 
in $p_{\perp}$-ordered terms, while retaining their strong points. 

The main trick that will be employed is to pick formal definitions of 
$p_{\perp}$, that simply and unambiguously can be translated into 
the older virtuality variables, e.g.\ for standard matrix-element 
merging. These definitions are based on lightcone kinematics, wherein 
a timelike branching into two massless daughters corresponds to 
$p_{\perp}^2 = z(1-z)M^2$ and the branching of a massless mother into 
a spacelike and a massless daughter to $p_{\perp}^2 = (1-z)Q^2$.
The actual $p_{\perp}$ of a branching will be different, and e.g.\
depend on the subsequent shower history, but should normally not 
deviate by much.

\section{TIMELIKE SHOWERS}

The new timelike algorithm is a hybrid between the traditional 
parton-shower and dipole-emission approaches, in the sense that 
the branching process is associated with the evolution of a single 
parton, like in a shower, but recoil effects occur inside dipoles. 
That is, a dipole partner is assigned for each branching, and energy 
and momentum is `borrowed' from this partner to give mass to the 
parton about to branch, while preserving the invariant mass of the 
dipole. (Thus four-momentum is not preserved locally for each
parton branching $a \to b c$. It was in the old algorithm, where
the kinematics of a branching was not constructed before the
off- or on-shell daughter masses had been found.)
Often the two partners are colour-connected, i.e.\ the 
colour of one matches the anticolour of the other, as defined by
the preceding showering history, but this need not be the case. 
In particular, intermediate resonances normally have masses that
should be preserved by the shower, e.g., in 
$\mathrm{t}\to\mathrm{b}\mathrm{W}^+$ the $\mathrm{W}^+$ 
takes the recoil when the $\mathrm{b}$ radiates a gluon.

The evolution variable is approximately the $p_{\perp}^2$ of a branching,
where $p_{\perp}$ is the transverse momentum for each of the two daughters 
with respect to the direction of the mother, in the rest frame of the 
dipole. (The recoiling dipole partner does not obtain any $p_{\perp}$ 
kick in this frame; only its longitudinal momentum is affected.) For 
the simple case of massless radiating partons and small virtualities 
relative to the kinematically possible ones, and in the limit that 
recoil effects from further emissions can be neglected, it agrees with
the $d_{ij}$ $p_{\perp}$-clustering distance defined in the 
\texttt{PYCLUS} algorithm \cite{Moretti:1998qx}.

All emissions are ordered in a single sequence $p_{\perp\mathrm{max}} > 
p_{\perp 1} > p_{\perp 2} > \ldots > p_{\perp\mathrm{min}}$. That is, 
each initial parton is evolved from the input $p_{\perp\mathrm{max}}$ 
scale downwards, and a hypothetical  branching $p_{\perp}$ is thereby 
found for it. The one with the largest $p_{\perp}$ is chosen to undergo 
the first actual branching. Thereafter, all partons now existing are 
evolved downwards from $p_{\perp 1}$, and a $p_{\perp 2}$ is chosen, 
and so on, until $p_{\perp\mathrm{min}}$ is reached. (Technically, the 
$p_{\perp}$ values for partons not directly or indirectly affected by 
a branching need not be reselected.) The evolution of a gluon is  
split in evolution on two separate sides, with half the branching
kernel each, but with different kinematical constraints since the
two dipoles have different masses. The evolution of a quark is also
split, into one $p_{\perp}$ scale for gluon emission and one for photon 
one, in general corresponding to different dipoles. 

With the choices above, the evolution factorizes. That is, a set of 
successive calls, where the $p_{\perp\mathrm{min}}$ of one call becomes 
the $p_{\perp\mathrm{max}}$ of the next, gives the same result (on the 
average) as one single call for the full $p_{\perp}$ range. This is the 
key element to allow Sudakovs to be conveniently obtained from trial 
showers \cite{Lonnblad:2001iq}, and to veto emissions above some 
$p_{\perp}$ scale, as required to combine different $n$-parton 
configurations efficiently.

The formal $p_{\perp}$ definition is
$p_{\perp\mathrm{evol}}^2 = z(1-z)(M^2 - m_0^2)$,
where $p_{\perp\mathrm{evol}}$ is the evolution variable, $z$ gives the 
energy sharing in the branching, as selected from the branching 
kernels, $M$ is the off-shell mass of the branching parton and 
$m_0$ its on-shell value. This $p_{\perp\mathrm{evol}}$ is also used as
$\alpha_{\mathrm{s}}$ scale.

When a $p_{\perp\mathrm{evol}}$ has been selected, this is translated 
to a $M^2 = m_0^2 + p_{\perp\mathrm{evol}}^2/(z(1-z))$. Note that the 
Jacobian factor is trivial:
$\mathrm{d}M^2/(M^2 - m_0^2) \; \mathrm{d}z = \mathrm{d}%
p_{\perp\mathrm{evol}}^2/p_{\perp\mathrm{evol}}^2 \; \mathrm{d}z$.  
From there on, the three-body kinematics of a 
branching is  constructed as in the old routine. This includes the 
detailed interpretation of $z$ and the related handling of nonzero 
on-shell masses for branching and recoiling partons, which leads to the 
physical $p_{\perp}$ not agreeing with the $p_{\perp\mathrm{evol}}$ 
defined here. In this sense, $p_{\perp\mathrm{evol}}$ becomes a formal 
variable, while $M$ really is a well-defined mass of a parton.

Also the corrections to $b\to b\mathrm{g}$ branchings ($b$ being a 
generic coloured particle) by merging with first-order 
$a\to bc\mathrm{g}$ matrix elements closely follows the existing 
machinery \cite{Norrbin:2000uu}, once the $p_{\perp\mathrm{evol}}$ has 
been converted to a mass of the branching parton. In general, the other 
parton $c$ used to define the matrix element need not be the same as 
the recoiling partner. To illustrate, consider a 
$\mathrm{Z}^0 \to \mathrm{q}\overline{\mathrm{q}}$ decay. Say the 
$\mathrm{q}$ branches first, $\mathrm{q} \to \mathrm{q}\mathrm{g}_1$. 
Obviously the $\overline{\mathrm{q}}$ then takes the recoil, and the new 
$\mathrm{q}$, $\mathrm{g}_1$ and $\overline{\mathrm{q}}$ momenta are used 
to match to the $\mathrm{Z}^0 \to \mathrm{q}\overline{\mathrm{q}}\mathrm{g}$ 
matrix element. The next time $\mathrm{q}$ branches, 
$\mathrm{q} \to \mathrm{q}\mathrm{g}_2$, the recoil is taken by 
the colour-connected $\mathrm{g}_1$ gluon, but the matrix element 
corrections are based on the newly created $\mathrm{q}$ and $\mathrm{g}_2$ 
momenta together with the $\overline{\mathrm{q}}$ (not the 
$\mathrm{g}_1$!) momentum. That way one may expect to achieve the most 
realistic description of mass effects in the collinear and soft regions.  

The shower inherits some further elements from the old algorithm, such as 
azimuthal anisotropies in gluon branchings from polarization effects.

The relevant parameters will have to be retuned, since the shower is 
quite different from the old mass-ordered one. In particular, it appears 
that the five-flavour $\Lambda_{\mathrm{QCD}}$ value has to be reduced 
relative to the current default, roughly by a factor of two (from 0.29 
to 0.14~GeV). 

\section{SPACELIKE SHOWERS}

Initial-state showers are constructed by backwards evolution
\cite{Sjostrand:1985xi}, starting at the hard interaction and 
successively reconstructing preceding branchings. To simplify 
the merging with first-order matrix elements, $z$ is defined by 
the ratio of $\hat{s}$ before and after an emission. For a massless 
parton branching into one spacelike with virtuality $Q^2$ and 
one with mass $m$, this gives
$p_{\perp}^2 = Q^2 - z (\hat{s} + Q^2)(Q^2 + m^2)/\hat{s}$, or
$p_{\perp}^2 = (1-z) Q^2 - z Q^4/\hat{s}$ for $m=0$.
Here $\hat{s}$ is the squared invariant mass after the emission, 
i.e.\ excluding the emitted on-mass-shell parton. 

The last term, $z Q^4/\hat{s}$, while normally expected to be small, 
gives a nontrivial relationship between $p_{\perp}^2$ and $Q^2$, 
e.g.\ with two possible $Q^2$  solutions for a given $p_{\perp}^2$.
To avoid the resulting technical problems, the evolution variable 
is picked to be $p_{\perp\mathrm{evol}}^2 = (1-z) Q^2$. Also here 
$p_{\perp\mathrm{evol}}$ sets the scale for the running 
$\alpha_{\mathrm{s}}$. Once selected, the $p_{\perp\mathrm{evol}}^2$
is translated into an actual $Q^2$ by the inverse relation
$Q^2 = p_{\perp\mathrm{evol}}^2/(1-z)$, with trivial Jacobian:
$\mathrm{d}Q^2/Q^2 \; \mathrm{d}z = \mathrm{d}%
p_{\perp\mathrm{evol}}^2/p_{\perp\mathrm{evol}}^2 \; \mathrm{d}z$.
From $Q^2$ the correct $p_{\perp}^2$, including the $z Q^4/\hat{s}$
term, can be constructed.

Emissions on the two incoming sides are interspersed to form a single
falling $p_{\perp}$ sequence, $p_{\perp\mathrm{max}} > 
p_{\perp 1} > p_{\perp 2} > \ldots > p_{\perp\mathrm{min}}$. 
That is, the $p_{\perp}$ of the latest branching considered sets 
the starting scale of the downwards evolution on both sides, 
with the next branching occurring at the side that gives the 
largest such evolved $p_{\perp}$. 

In a branching $a \to b c$, the newly reconstructed mother $a$ is 
assumed to have vanishing mass --- a heavy quark would have to be 
virtual to exist inside a proton, so it makes no sense to put it on 
mass shell. The previous mother $b$, which used to be massless, now 
acquires the spacelike virtuality $Q^2$ and the correct $p_{\perp}$ 
previously mentioned, and kinematics has to be adjusted accordingly. 

In the old algorithm, the $b$ kinematics was not constructed until 
its spacelike virtuality had been set, and so four-momentum was 
explicitly conserved at each shower branching. In the new algorithm,
this is no longer the case. (A corresponding change occurs between 
the old and new timelike showers, as noted above.) Instead it is the 
set of partons produced by this mother $b$ and the current mother $d$ 
on the other side of the event that collectively acquire the 
$p_{\perp}$ of the new $a \to b c$ branching. Explicitly, when the 
$b$ is pushed off-shell, the $d$ four-momentum is modified accordingly, 
such that their invariant mass is retained. Thereafter a set of 
rotations and boosts of the whole $b+d$-produced system bring them 
to the frame where $b$ has the desired $p_{\perp}$ and $d$ is restored 
to its correct four-momentum. 

Matrix-element corrections can be applied to the first, i.e.\ hardest
in $p_{\perp}$, branching on both sides of the event, to improve the 
accuracy of the high-$p_{\perp}$ description. Also several other aspects 
are directly inherited from the old algorithm. 

Work on the algorithm is ongoing. In particular, an optimal description
of kinematics for massive quarks in the shower, i.e.\ $\mathrm{c}$ and 
$\mathrm{b}$ quarks, remains to be worked out.

Some first tests of the algorithm are reported elsewhere 
\cite{showerpthere}. In general, its behaviour appears rather similar
to that of the old algorithm.

\section{OUTLOOK} 

The algorithms introduced above are still in a development stage. 
In particular, it remains to combine the two. One possibility would be to 
construct the spacelike shower first, thereby providing a list of 
emitted partons with their respective emission $p_{\perp}$ scales. 
This list would then be used as input for the timelike shower, where 
each emission $p_{\perp}$ sets the upper evolution scale of the 
respective parton. This is straightforward, but does not allow a fully 
factorized evolution, i.e.\ it is not feasible to stop the evolution 
at some $p_{\perp}$ value and continue downwards from there in a 
subsequent call. The alternative would be to intersperse spacelike 
and timelike branchings, in one common $p_{\perp}$-ordered sequence.

Obviously the finished algorithms have to be compared with data,
to understand how well they do. One should not expect any major 
upheavals, since checks show that they perform similarly 
to the old ones at current energies, but the hope is for a somewhat
improved and more consistent description. The step thereafter would 
be to study specific processes, such as $\mathrm{W} + n$ jets, to find 
how good a matching can be obtained between the different $n$-jet 
multiplicities, when initial parton configurations are classified by 
their $p_{\perp}$-clustering properties. The \texttt{PYCLUS} algorithm 
here needs to be extended to cluster also beam jets. Since one cannot 
expect a perfect match between generated and clustering-reconstructed 
shower histories, it may become necessary to allow trial showers and 
vetoed showers over some $p_{\perp}$ matching range, but hopefully then 
a rather small one. If successful, one may expect these new algorithms 
to become standard tools for LHC physics studies in the years to come. 

\bibliography{biblifile}

\end{document}